\documentclass[journal=jacsat,manuscript=article]{achemso}

\usepackage[version=3]{mhchem} 

\author{Mateusz Chwastyk}
\affiliation{Institute of Physics, Polish Academy of Sciences, Al. Lotnik\'ow 
32/46, PL-02668 Warsaw, Poland}
\email{chwastyk@ifpan.edu.pl}
\author{Marek Cieplak}
\affiliation{Institute of Physics, Polish Academy of Sciences, Al. Lotnik\'ow 
32/46, PL-02668 Warsaw, Poland}

\title{Conformational Biases of $\alpha$-synuclein and Formation of Transient Knots}

\begin{document}

\maketitle

\let\thefootnote\relax\footnotetext{$^{*}$ corresponding author}






\begin{abstract}
\noindent
We study local conformational biases in the dynamics of $\alpha$-synuclein
by using all-atom simulations with explicit and implicit solvents. 
The biases are related to the frequency of the specific contact formation.
In both approaches, the protein is intrinsically disordered 
and its strongest bias is to make bend, and then turn, local structures.
The explicit-solvent conformations can be substantially more extended 
which allows for formation of transient trefoil knots, both deep
and shallow, that may last for up to 5 $\mu$s.
The two-chain self-association events, both short- and long-lived, are
dominated by formation of contacts in the central part of the 
sequence. This part tends to form helices when bound to a micelle.

\end{abstract}

\maketitle

\section{Introduction}

$\alpha$-Synuclein is a 140-residue protein that is found in the mammalian
brain as both a soluble and membrane-associated molecule \cite{Mesliah}.
It is highly expressed in the mitochondria and is
active in the presynaptic region, where it is involved in
the regulation of synaptic vesicle release, vesicle trafficking, and
fatty-acid binding \cite{Melki}. 
It is also a chaperon protein that affects
the degradation and assembly of the presynaptic SNARE complex \cite{Mesliah}.
$\alpha$-Synuclein has a propensity to form self-propagating fibrils that
are present in the brain of most of the Parkinson-disease patients \cite{Stefanis}.
It is unclear whether
toxicity arises as a consequence of the fibers
themselves or of the presence of oligomeric intermediates, which appear
en route to fiber formation \cite{Melki}.\\

The structural properties of the monomeric $\alpha$-synuclein are
then of great interest and also a matter of uncertainty.
When bound to a micelle, the protein is partially structured
(the structure code PDB:1XQ8), as evidenced by NMR studies \cite{Ulmer}:
the segments 2-37 (denoted here as $h_1$) and 45-92 ($h_2)$ 
are $\alpha$-helical. They form a hairpin
linked by a loop between the helices. The N-terminal region is
charged as it is lysine-rich. This feature enables interactions
with a membrane \cite{Mesliah} and makes the region disordered.
However, in the absence of binding to the micelle,
both the tryptophan fluorescence \cite{Rooijen}
and NMR studies \cite{Bax} indicate the existence of intrinsic disorder 
throughout the monomeric sequence. \\

$\alpha$-Synuclein has also been studied by molecular dynamics simulations.
Several of them \cite{Gordon,Scheraga,Briels}, done within all-atom
explicit solvent schemes, lead to the conclusion that $\alpha$-synuclein is
indeed disordered since the simulations reveal the
existence of rapid temporal changes in the local conformational
biases. The biases can be determined through the DSSP scheme \cite{dssp}.
In particular, the occurrence of the
$\beta $ structures in single chains is low \cite{Briels} which does
not favor dimeric aggregation.\\

Here, we study the free protein by all-atom simulations of two kinds: with
the explicit and implicit solvents. 
We determine the time-averaged probabilities for a residue $i$
to belong to a particular secondary structure and demonstrate
that both approaches indicate the presence of the all-sequence disorder.
These probabilities are found to be the largest
for the bend and turn local conformations whereas the helical
ones rank the third.\\

The very sequential pattern of the probabilities is found to depend on the
nature of the solvent. More importantly, we demonstrate that the interactions 
with the molecules of water generally introduce a qualitative difference to the
physics of $\alpha$-synuclein: the explicit solvent conformations tend
to be  substantially more extended. Such conformations
allow for much more frequent global opening and closing events,
resulting in higher probability to form knotted backbones.
In the implicit-solvent case, we observe no knots.
In contrast, in one trajectory calculated with the explicit solvent
we observe a deeply knotted trefoil topology that lasts for about 5 $\mu$s.
In another trajectory, we observe a series of shallow knots that
emerge and disappear within a similar time span.\\

While the presence of knots in many structured proteins is 
well established by now \cite{Faisca,Jackson, Sulkowska,KnotProt},
this is not so in the disordered case.
One difficulty is related to the fact that the disordered
conformations are not readily available and have to be generated
through computer simulations. Another is that the knotted structures
are only transient -- in our simulations, they last for of order 5 $\mu$s.
However, as we discuss at the end of the paper, even then they may impede
proteasomal degradation and thus enhance accumulation of the proteins.
We expect that other intrinsically disorder proteins may also tangle
to the knotted forms, but the corresponding characteristic time scales
for their stability should be protein dependent.\\

The propensity to form transient local structures is related to the temporary
establishment of inter-residue interactions that can be described,
in the coarse-grained sense, as the establishment of inter-residue contacts. 
Our definition of a contact is based on the existence
of overlaps between effective spheres associated with the heavy atoms
belonging to separate residues (see the Methods section).
We find that the most likely contacts are those that involve the N-terminus
(e.g. by linking to the $h_2$ region) and those that occur within the $h_1$ region. 
The contacts in the latter region do not lead to any persistent helix.\\

We then analyze the self-association of two disordered $\alpha$-synuclein chains 
and find that the most frequent contacts are those that connect
$h_2$ in one chain with $h_2$ in another chain. The next in importance
are those that connect $h_1$ with $h_2$. 
These findings are consistent with the expectation that
the capacity for association is related to the central region 65-90.
The absence of this region has been shown experimentally to impede 
oligomerization and fibrilogenesis \cite{Mesliah}. Interestingly,
we also find that the dimeric inter-chain contacts are distinct
from those that are likely to arise during the single-chain dynamics.
The involvement of the $h_1$ and $h_2$ regions, that are helical when
bound to the micelle, seems to echo the results of studies
\cite{Scheraga,Thompson1,Thompson2} pertaining to the tetrameric aggregation
which appear to lead to the emergence of a ring of helices (especially
when the partially structured PDB:1XQR conformation is used to generate the
starting states). Such ring states resist any further aggregation \cite{Bax} and
may serve to store excessive $\alpha$-synucleins
in the cell\cite{Pochapsky}.\\

\section{Methods}

Our implicit-solvent all-atom simulations were conducted
using the NAMD code version CVS-2013-11-07 for
Linux-x86\_64-MPI\cite{namd}
with the CHARMM36n force field  \cite{CHARMM36n} that is refined to be
applicable to the intrinsically disordered proteins.
The snapshots of the conformations were plotted by using the
VMD package \cite{vmd}.
The simulations were performed with the use of the Generalized Born
Implicit Solvent method \cite{Born}.
In the single-chain simulations, we used free boundary conditions.
The cut-off radius for non-bonded interactions
was set to 1.4 nm.  We have generated 5 trajectories,
each lasting for 30 ns.
The first part of the simulations involved energy minimization of the system for
5 000 conjugate gradient steps. In the second part, the system was heated
up to $T$ of 298 K.
The temperature  was controlled by the standard Langevin algorithm.
The time step was 1 fs.
The conformations of the protein were captured every 2 ps.
The 5 starting conformations were obtained by random selection from 5 different 
points of the ANTON-derived trajectory.\\


When studying the association processes of multiple chains, we switch to periodic boundary
conditions and use the smooth Particle Mesh Ewald procedure \cite{essmann}
to treat the long range electrostatics.
The procedure is a variant of the Particle Mesh Ewald method \cite{Darden}
as implemented in NAMD. The grid spacing is 0.16 nm.
In the simulation of two chains, we used the cubic simulation box
with the edge length of  60.0 nm. We prepared 
50 starting systems in the following way.
We first chose all possible pairs from the set 
of the initial structures that were simulated as single chains.
This yields 10 different sets of pairs. For each of them,
we prepared 5 different conformations by
changing the mutual orientations 
of the chains by using the PyMol software \cite{Pymol}. Each starting
system was simulated for 30 ns which yielded 15 000 frames.
Altogether, we have considered 750 000 conformations.\\

Independent of the number of chains, the conformations can be
characterized by their instantaneous contact maps.
The contacts may link residues within chains and between chains.
We determine their presence by checking if there are
atomic overlaps between residues.
Specifically, each non-hydrogen heavy atom is represented by
a sphere of radius equal to the atom's van der Waals radius enlarged
by a factor of 1.24 to account for attractive interactions
(the factor corresponds to the inflection point in the Lennard-Jones potential).
The values of the van der Waals radii are taken from ref. \cite{Tsai}.
A contact between residues $i$ and $j$ is said to exist if 
the two residues contain at least two spheres, one belonging to
$i$ and the other to $j$, that overlap.
In the terminology of ref. (Wolek et al.)\cite{Wolek},
this procedure is denoted as OV. We have successfully used such
contact maps in the studies of protein stretching \cite{Sikora}.\\

\section{Results \& Discussion}

We have obtained the implicit-solvent results through five 30 ns
NAMD-based simulations \cite{namd}, as described in the Methods section. 
The explicit-solvent trajectories have been derived by Robustelli
et al. \cite{Robustelli} when developing a set of novel molecular force
fields that would be adequate for both structured and disordered proteins. 
$\alpha$-Synuclein was one of several proteins considered in 
ref. \cite{Robustelli} in the tests. These novel force fields
were designed to be used on the special purpose supercomputer ANTON.
For each of the considered force fields, $\alpha$-synuclein comes out as a fully
disordered protein. In particular, the local $\alpha$-helical content, though 
non-zero in certain parts of the sequence (e.g. around residue 20), has been
found never to exceed 45\% for any of the force fields used. 
Here, we analyze two trajectories, one 30-$\mu$s and the other 20-$\mu$s long, that
were obtained using the C22* force field and the TIP4P-D water molecules.\\

An earlier extensive explicit solvent study by Sethi et al. \cite{Gnanakaran}
involved  GROMACS \cite{GROMACS}, the OPLS-AA force field \cite{OPLS},
and a division of the protein into weakly interacting modules.
Our results on the propensities to form helices and $\beta$-bridges
are consistent with this study (when the starting conformations
are random). However, the novelty of our research in this context,
also compared to that of ref. \cite{Robustelli}, is showing that the dominant
structural bias is for the T and S secondary structures and that the
structural biases can be explained in terms of contact maps.\\


{\bf The transient secondary structures}\\

Here, we consider the probabilities, $P_{\lambda}$, of forming secondary structures,
$\lambda$, at the sequential residue $i$. In addition to the turn (T) and bend (S), the
secondary structures detected are helix (H), $\beta$-bridge (B), and helix 3$_{10}$ (G).
We find no appreciable probabilities for strands (E) and helices-$\pi$ (I).
When there is no detectable structure at a residue, we refer to it as a
coil state (C).\\

\begin{figure}[h]
\centering
\includegraphics[width=0.8\textwidth]{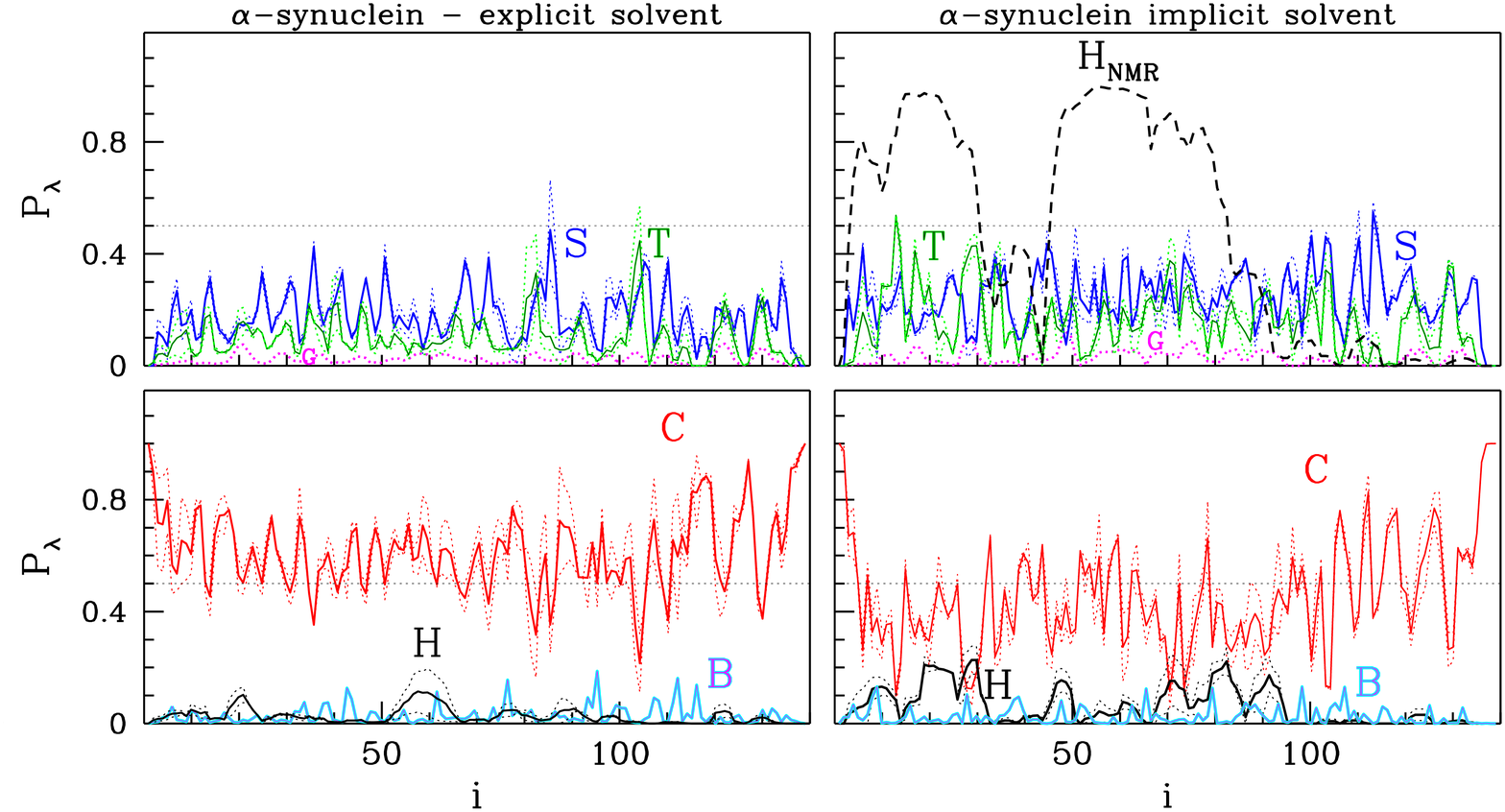}
\caption{Probabilities $P_{\lambda}$ for the monomer of $\alpha$-synuclein to adopt the local
secondary structures $\lambda$ at residue $i$. The structures shown are T, S
(the upper panels, green and blue respectively) and
H, B, G, and C (the lower panels, black, purple, magenta, and red respectively).
The dotted lines (not shown for G and B for clarity of the presentation) indicate
the size of the error bars. They were obtained by splitting the whole trajectory
into two halves. The data points in the left panels have been obtained by using the
C22*/TIP4P-4D force field with the explicit solvent.
The panels on the right correspond to the NAMD-derived implicit solvent simulations.
The black broken line shows the helical
content if the starting conformations is the PDB:1XQ8 structure.
} \label{synan}
\end{figure}

The time-averaged data shown in the left panels of Figure \ref{synan} is based
on the explicit-solvent 30 $\mu$s simulation performed on the
special purpose ANTON supercomputer
at D.E.Shaw Research in New York \cite{Robustelli}.
The $P_H(i)$ function for the C22*/TIP4P-4D force field never exceeds
20\%. Neither does $P_B(i)$. $P_G(i)$ does not exceed 10\%.
It is seen that the dominant tendency is to form the local S and T
structures, sometimes in the same segment. The system is clearly disordered
because the various structural propensities switch in time and compete
at essentially all residue. At about 8\% of the residues, $P_C(i)$ is under
50\% (the level indicated by the dotted horizontal line) but at all
remaining residues the lack of any detectable order is dominant behavior.\\

The right panels of Figure \ref{synan} show similar data for the all-atom $5 \times$ 30 ns
implicit-solvent simulation. They are qualitatively similar in that
the strongest structural propensities are for the S and T local arrangements.
However, the overall propensities for making secondary structures are stronger
than those shown in Figure \ref{synan}. In particular, at most residues, the
propensity to show no structure (C) is smaller than 50\%.
In most cases, the starting conformations were chosen randomly from the ANTON-derived 
trajectory. In this way, the two kinds of simulations are expected to be
in similar regions of the phase space. Thus we conclude that the larger
structural propensities are due to the lack of the molecules of water.
In order to quantify the differences between the explicit and implicit solvent
results, we residue average $P_{\lambda}$ along the sequence. For the explicit
solvent case, we obtain 0.026 (H), 0.096 (T), 0.179 (S), 0.028 (B), 0.022 (G),
0.000 (I), 0.033 (E), and 0.614 (C) where the letters in brackets indicate
the type of the secondary structure. For the implicit solvent case, we
obtain 0.062 (H), 0.161 (T), 0.243 (S), 0.025 (B), 0.029 (G), 0.000 (I),
0.025 (E), and 0.455 (C). The biggest shift is seen to be for the C-content.
Independent of the model, the largest ordering tendency is for S.\\

The top-right panel of Figure \ref{synan} also shows results for $P_H(i)$ for 5
trajectories that start from the PDB:1XQ8 conformation. They are seen to boost $P_H(i)$
substantially, indicating a strong sensitivity to the initial conditions
within the simulational times considered.\\

Even though the implicit-solvent simulations yield results that are
qualitatively similar to those of the explicit-solvent ones, other than
the general shift toward more ordering, there is a sequential displacement
in the patterns. When one plots the ratio
of $P_{\lambda(i)}$ obtained with the implicit solvent
to $P_{\lambda} (i)$ with the explicit solvent
then, for any $\lambda$,  
there is a significant patterning as a function of $i$ (not shown).\\

\begin{figure}[h]
\centering
\includegraphics[width=0.5\textwidth]{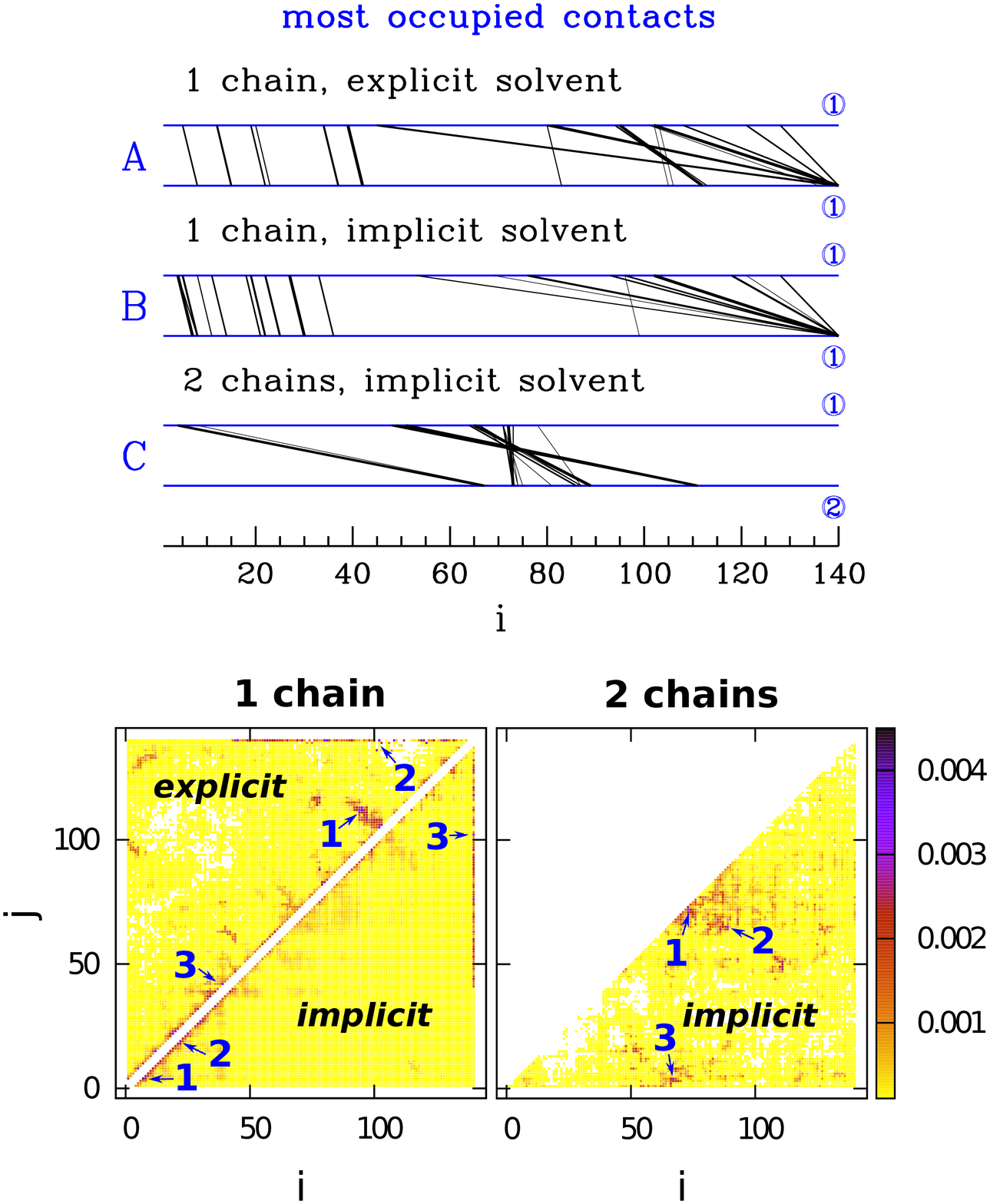}
\caption{The top three panels: The 20 most likely contacts that participate the the $\alpha$-synuclein
dynamics. Panels A, B and C correspond to: the single-chain dynamics with
the explicit solvent, single-chain dynamics with the implicit solvent and
the two-chain dynamics with the implicit solvent respectively. 
The horizontal lines indicate
locations along the sequence. Contacts link locations in the upper
line with the locations in the lower line: contacts are either within
the same chain (the top part) or with another chain (the bottom part).
The higher the ranking, the thicker the line.
The occupations are counted in the cumulative fashion throughout the
simulation. We show contacts corresponding to 
locations $i$ and $j$ where $i<j$. For simplicity, we assume the
symmetry between $i$ and $j$.
In panel A, the most highly occupied contacts is 95-112. The other contacts
shown are 102-140 (rank 2), 39-42 (3), 81-140 (4), 12-15 (5),
80-140 (6), 34-37 (7), 45-140 (8), 121-139 (9), 19-22 (10), 94-112 (11),
108-140 (12), 5-8 (13), 128-140 (14), 95-113 (15), 20-23 (16),
80-83 (17), 102-105 (18), 101-136 (19), and 103-106 (20).
The total number of contacts identified was 8606.
In panel B, the most highly occupied contact is 4-7.
The other contacts shown are: 27-30 (rank 2), 102-140 (3), 19-22 (4), 
22-25 (5), 76-140 (6), 118-140 (7), 5-8 (8), 128-140 (9), 18-21 (10),
11-14 (11), 96-140 (12), 33-36 (13), 93-140 (14), 53-140 (15), 8-11 (16),
4-8 (17), 69-140 (18), 96-99 (19), and 121-140 (20).
The total number of contacts identified was 9454.
In panel C, the most highly occupied interchain contact is 72-73.
The other contacts shown are: 65-89 (rank 2),
4-67 (3), 51-111 (4), 49-111 (5), 48-111 (6), 64-86 (7),  71-73 (8),
73-73 (9), 66-87 (10), 8-68 (11), 49-110 (12), 78-87 (13), 4-66 (14),
50-111 (15), 65-87 (16), 71-75 (17), 72-74 (18), 71-74 (19), and 66-81 (20).
The total number of contacts identified was 8596.
The bottom panel: The time averaged contact map corresponding to the
situations A, B, and C. The color code, with the scale on the right,
indicates the probability of the occurrence of a contact. The white spots
corresponds to regions in which no contacts were found. The contact maps 
are symmetric with respect to the diagonal so only a half of any map is shown.
The digits indicate the rankings of the top three contacts.
} \label{aggr}
\end{figure}

Panels A and B of Figure \ref{aggr} show 20 of the most likely
contacts that form during the single-chain evolution. 
The panels are for the explicit and implicit solvent cases respectively.
There is a qualitative similarity in their looks but the differences
in the details of the pattern are also evident.
There are four contacts appear in both panels: 
5-8, 19-22, 102-140, and 1-140. Furthermore, contact 121-139 
in panel A is almost identical to contact 121-140 in panel B.
However, 15 other contacts are distinct. Despite the differences,
all of these top contacts are seen to 
be related to the attempts to construct the first helix and to 
connect the N-terminal part with the center region.
The increased number of strong contacts in the N-terminal segment
extending up to residue 36 explains the larger tendencies to form
helices in the implicit-solvent case.
The bottom-left panel of Figure \ref{aggr} show the probability
density corresponding to all contacts that have been detected
in the single-chain simulations. The contact maps corresponding to the
implicit- and explicit-solvent simulations are seen to be distinct.\\



{\bf The geometrical characteristics of conformations}\\

\begin{figure}[h]
\centering
\includegraphics[width=0.8\textwidth]{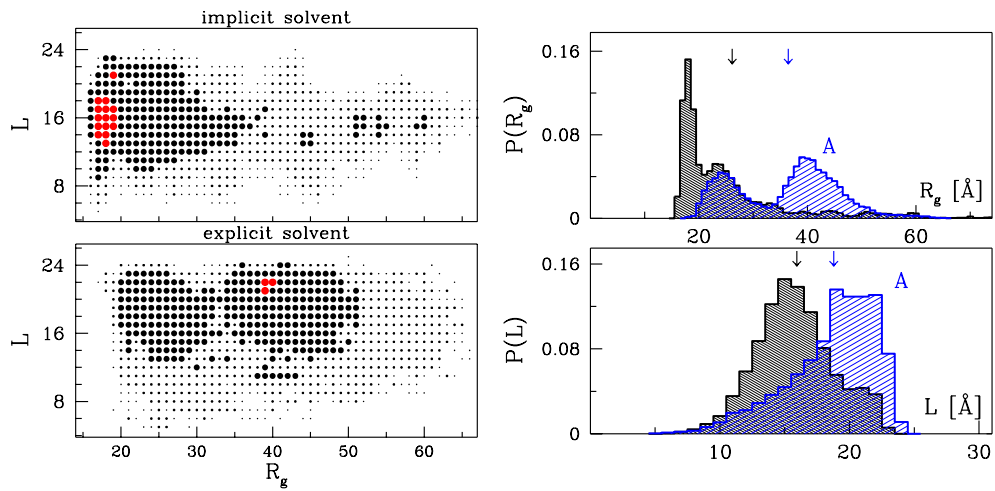}
\caption{The left panels: The $R_g$-$L$ cross plot for $\alpha$-synuclein.
The data points have been obtained
by summing the values belonging to 1 {\AA} $\times$ 1 {\AA} squares and showing
only the squares with the probability of occupation that exceeds a threshold.
For the red data points, the threshold is 0.01. For the black data points it is
0.001, 0.0001, and 0.00001, in the diminishing order of the size of the symbols.
The right panels: The probability distributions of $R_g$ (the top panel) and $L$ (the
bottom panel). The histogram in black is for the implicit-solvent data.
The histogram in blue, also denoted by A, is for the explicit-solvent case.
The bin size is 1 {\AA}.
The arrows indicate the corresponding mean values.
} \label{greecom}
\end{figure}

A qualitative way to asses the free-energy landscape is by plotting the
visited points on the $R_g$-$L$ plane, where $R_g$ denotes the radius of
gyration and $L$ the end-to-end distance. The regions that are the 
most frequently visited are expected to correspond to the lowest free energy.
The left panels of Figure \ref{greecom} shows such plots for the implicit- and
explicit-solvent systems considered (top and bottom respectively). 
The landscapes are seen to have substantial regions
that overlap but the highest frequency regions (marked in red) are shifted 
significantly between the two descriptions. This is seen especially
in the values of $R_g$: the explicit-solvent conformations
are much less globular. This feature is also reflected in the distributions
of $R_g$ and $L$ shown in the right panels of Figure \ref{greecom}. 
The distribution of $R_g$
is double peaked for both systems but the maxima of the peaks are shifted
upward in the explicit-solvent case. The distribution of $L$ is single-peaked
but the peak is also upward-moved.\\

The top-right panel of Figure \ref{greecom} allows for making comparisons with experiments.
Recent SAXS study of $\alpha$-synuclein obtained from human 
blood cells \cite{Araki} yields the average $R_g$ between 33.1 and 
33.3 {\AA} depending on the buffer used (pH=7.4). This is consistent with 
our explicit-solvent value of 36.2 {\AA} compared to 25.8 {\AA}
indicating that the explicit-solvent approach is closer to the reality.
It should be noted that the same SAXS study yields results
that depend strongly on the buffer (between 27.2 and 42.7 {\AA})
if one uses a recombinant
$\alpha$-synuclein. The authors attribute this fact to the
"harsh" treatment used in the preparatory work of such proteins.\\

{\bf Dynamics of the knot formation}\\

\begin{figure}[h]
\centering
\includegraphics[width=0.8\textwidth]{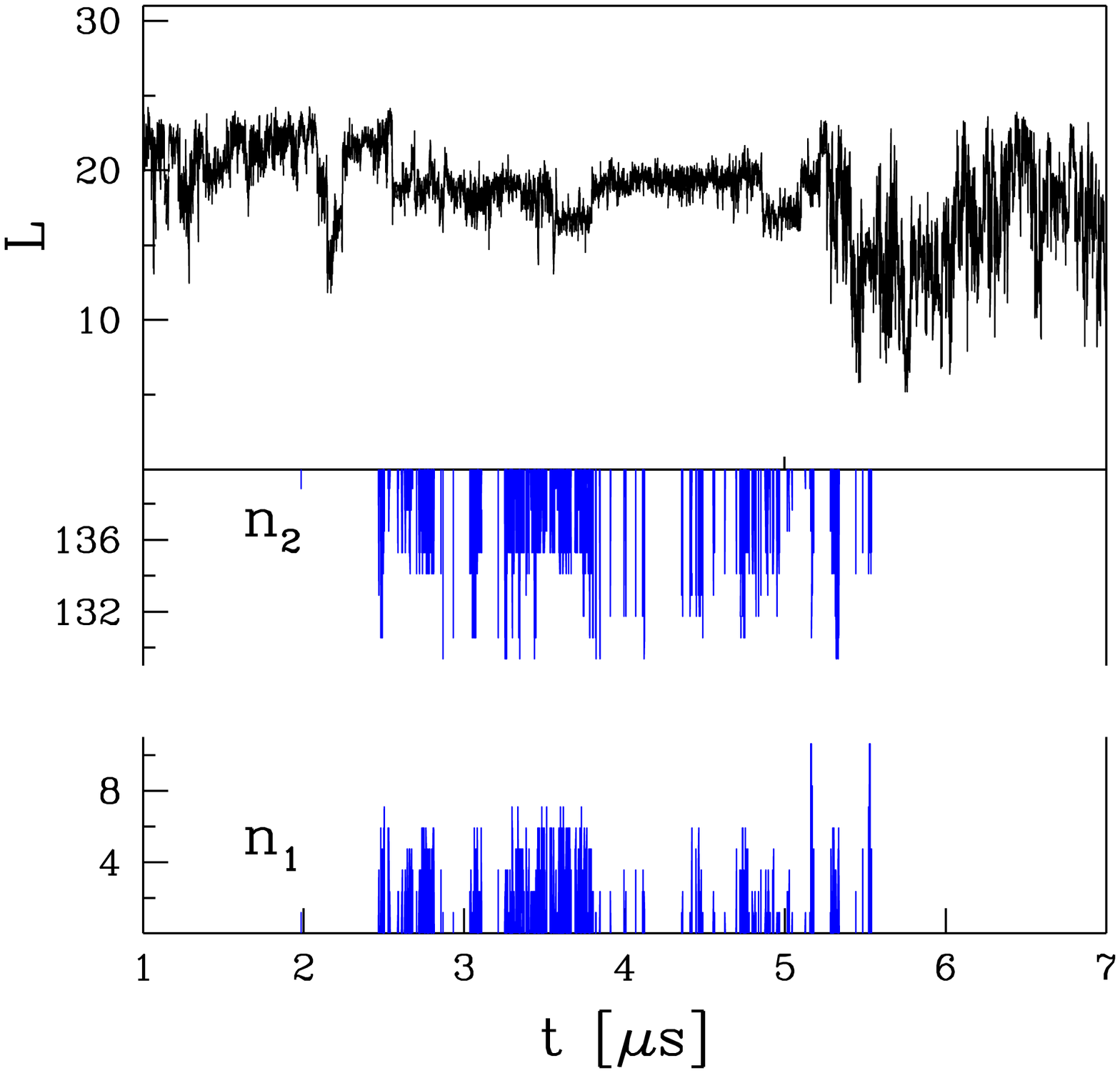}
\caption{A 6-$\mu$s fragment of the explicit solvent trajectory. The upper panel
shows the end-to-end distance. The lower panel shows the locations of the knot ends.
} \label{ends}
\end{figure}

We test the presence of knots by using the KMT algorithm \cite{KMT,KMT1}.
We have not spotted any knots in the implicit-solvent trajectories.
However, the much more expansive (on the $R_g$-$L$ plane) explicit solvent
trajectory is found to have a 3 $\mu$s long time segment in which a shallow 
trefoil knot keeps forming and disappearing very much like what happens
with the shallowly knotted proteins at the air-water interface \cite{Yani}.
This can be illustrated by showing the sequential locations of the knot ends
$n_1$ and $n_2$. The knot ends are determined by truncating the sequence
from both termini until coming to a stage at which the knot dissolves.
Figure \ref{ends} shows that the locations of $n_1$ and $n_2$ fluctuate
but their distances from the closest termini (N for $n_1$ and C for $n_2$)
never exceed the span of 10 residues and sometimes can be as short as 2 residues.
It should be noted that for most of the discussed segment of time, $L$
varies little and stays in the upper range of the values shown in
the left panels of Figure \ref{greecom}.\\

\begin{figure}[h]
\centering
\includegraphics[width=0.8\textwidth]{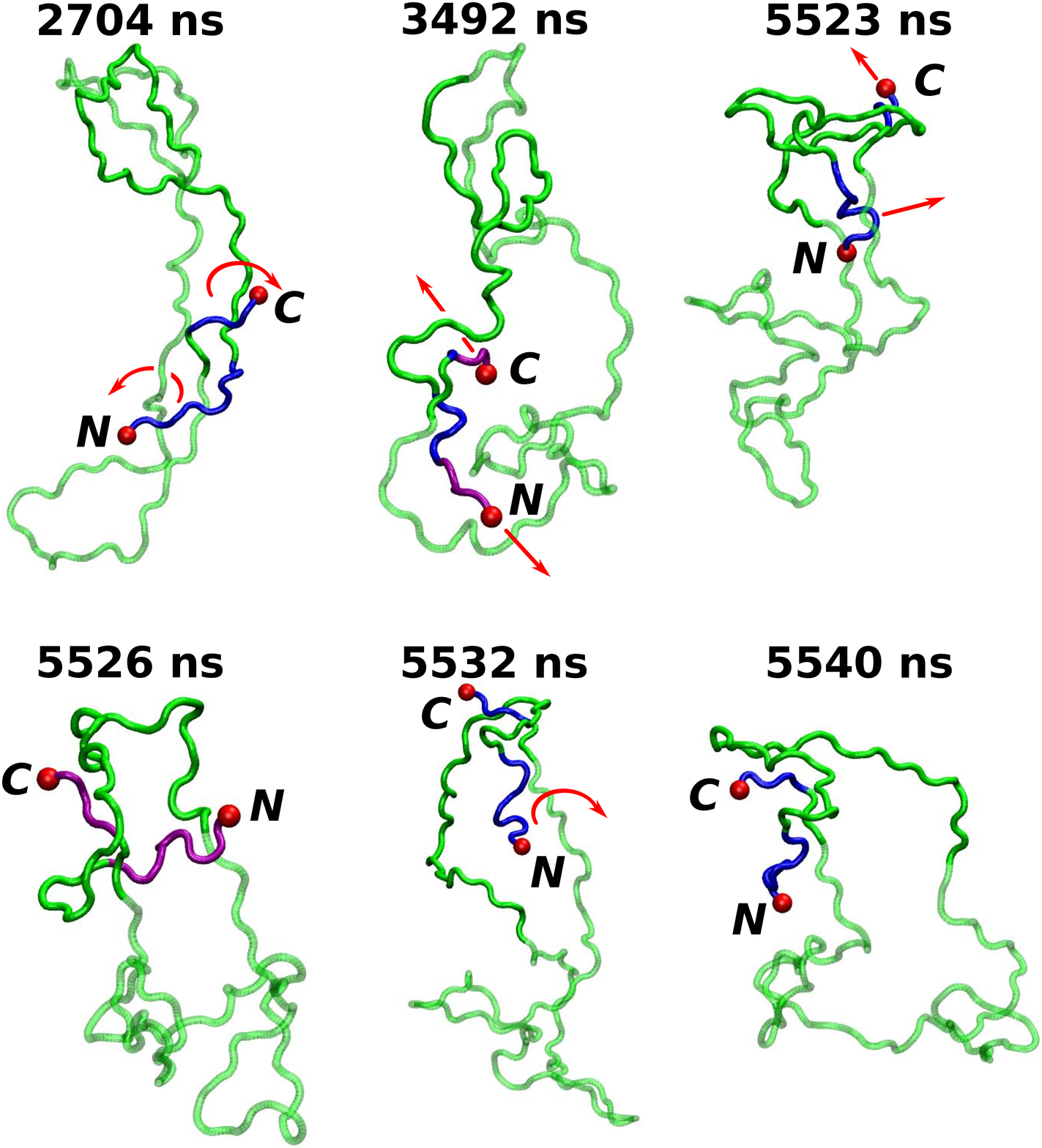}
\caption{Examples of conformations in the fragment of the trajectory shown
in Figure \ref{ends}. 
The conformation at 2704 is the state just before the knotting process begins.
The panels for 3492 and 5526 correspond to situations in 
which the knot exists. In the remaining panels, there is no knot.
In the panel corresponding to 5526 ns, the knotted segment is in
green and the purple lines indicate the segments between the
knot ends and the nearest termini. In this example, the purple segments
are the longest that were found. In the panel corresponding to 3492 ns the
sequence within the knot is longer: it combines the segments in green and blue.
The segments in purple are outside of the knots and are shorter than
the maximal ones (for 5526 ns). In the remaining (unknotted panels)
the segments in blue are equal in length to the maximal segments in purple
shown for 5526 ns. They merely indicate the regions to look at
when a knot is about to form. The conformation at 3492 is obtained
by performing moves indicated by the red arrows.
The last stage here corresponds to the direct threading. In the
panel for 5523 ns, a slipknot is formed and it further indicated motion
leads to knotting (5526 ns). Further unknotting takes place by
driving the slipknot out of the knot loop (5532 ns).
The conformation at 5540 represents the completely unknotted state. The
knot was no longer observed during this trajectory after this stage.
} \label{knots}
\end{figure}

Figure \ref{knots} shows an example of knot formation by direct 
threading (the panel corresponding to 3492) or slipknotting (5523-5526 ns)
and unknotting by the slipknot mechanism. A slipknot conformation
is one in which a bend segment of the backbone pierces
through a backbone (knot) loop, It transforms into a knot on straightening
the segment \cite{KnotProt}.\\

\begin{figure}[h]
\centering
\includegraphics[width=0.8\textwidth]{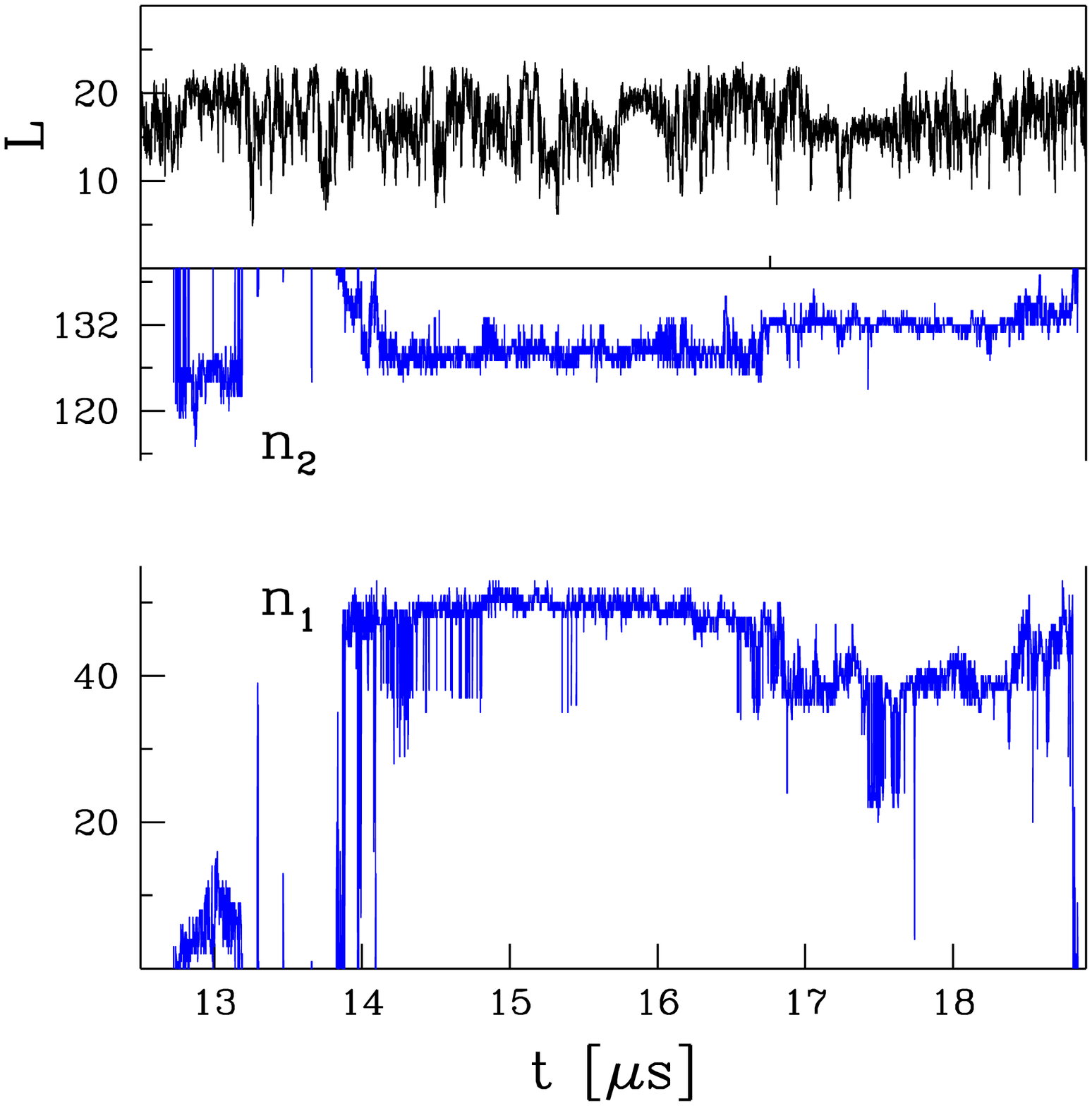}
\caption{A 6-$\mu$s fragment of the explicit solvent trajectory. The upper panel
shows the end-to-end distance. The lower panel shows the locations of the knot ends.
} \label{ends1}
\end{figure}

\begin{figure}[h]
\centering
\includegraphics[width=0.8\textwidth]{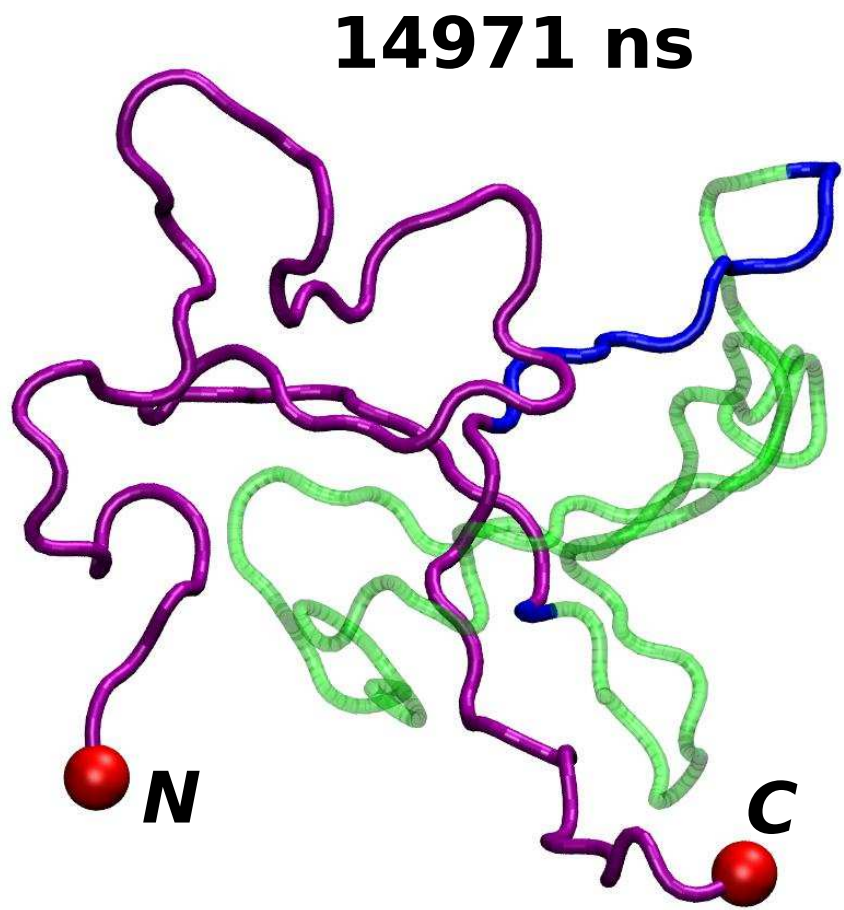}
\caption{A snapshot
of a knotted trefoil structure in which the sequential
distance between the knot ends is 73 -- the shortest
observed in the second trajectory.
The knotted segment is in green and blue and the knot ends
are at the boundary of blue and purple. These are residues
52 and 126. In other conformations,
the knot ends may move into the core of the knot along
the segments shown in blue but the separation between them
will become larger than the minimal value.
The motion of the ends along the purple segments eventually
leads to the dissolution of the knot.
} \label{tref}
\end{figure}

We have also considered a second available trajectory that is 20 $\mu$s long.
Figure \ref{ends1} shows that in nearly a quarter of the
duration of this trajectory there is a stable deeply knotted topology.
The knot ends of the corresponding conformations keep sliding to an
extent, as shown and explained in Figure \ref{tref}.\\

It should be pointed out that the knot formation in both trajectories
is kinetically driven and that the knots, though fairly long
lasting, are transient.\\

In an effort to assess the consistency of the knotted conformations from 
the a99SB-disp and c22*/TIP4P-D explicit solvent trajectories with previous 
experiments \cite{Schwalbe}, we back-calculated several NMR observables from the
knotted conformational ensembles and compared them to a suite of previously reported 
NMR measurements in Table \ref{table1}. These measurements include chemical 
shifts, residual dipolar couplings (RDCs) and backbone scalar coupling constants 
which report on local backbone conformations, as well as paramagnetic relaxation 
enhancements (PREs) which report on transient tertiary contacts.   
We found the knotted portions of the c22*/TIP4P-D and a99SB­-disp trajectories 
show slightly worse agreement with experimental measurements compared to their 
parent trajectories, but are in substantially better agreement than simulations 
run with similar force fields and the TIP3P and TIP3P-CHARMM water models.  
This comparison suggests that a sub-population of knotted conformations similar 
to those observed in these simulations would not be grossly inconsistent with 
pervious experimental measurements, though the populations may be smaller than 
those observed in these trajectories, particularly in the c22*/TIP4P-D trajectory.\\

{\bf Self-association in the implicit solvent approach}\\

\begin{figure}[h]
\centering
\includegraphics[width=0.8\textwidth]{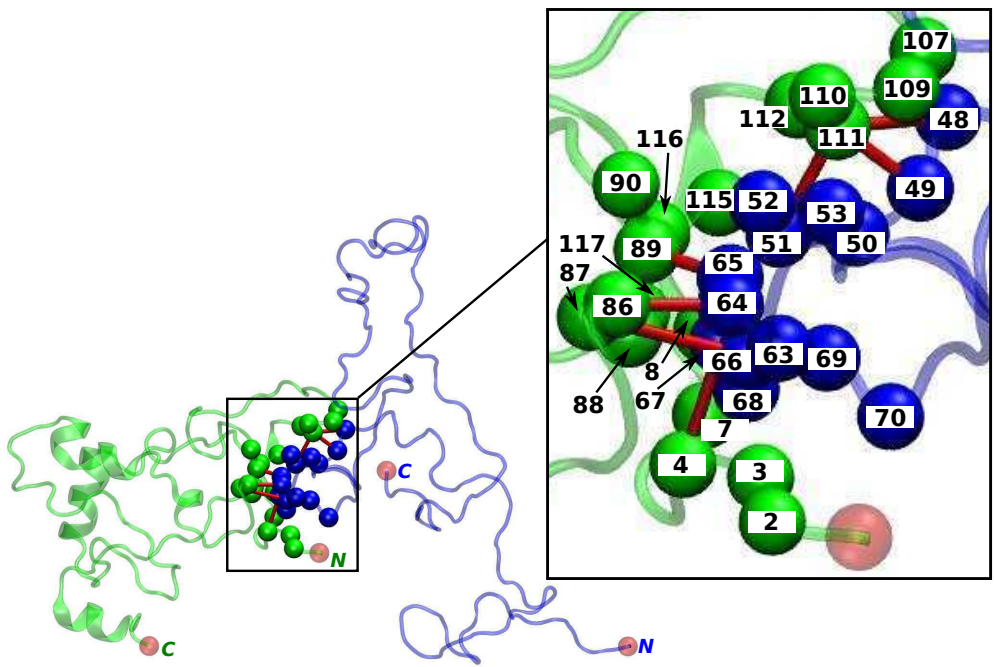}
\caption{An example of a two-chain associated state.
One chain is shown in the shades of blue and another in the
shades of green. The interfacial residues are shown as spherical beads,
in blue and green correspondingly. Other residues are not shown.
There are 40 inter-chain contacts.
Seven of these contacts are marked as red lines. They 
belong to the top-ten most probable contacts. These are: 65-89 (rank 2),
4-67 (3), 51-111 (4), 49-111 (5), 48-111 (6), 64-86 (7), and 66-87 (10)
(see also panel C in Figure \ref{aggr}).
The ranking is based on all association events. The event shown
is a part of the longest lasting dimer (see Figure \ref{times}).
In this particular dimer, the most frequent contact is 4-67 (marked in red)
and then 51-112. In the snapshot shown, 51-112 does form a contact but,
unlike 51-111, it is not marked because it does not belong to the
list of top 20 contacts derived from all events, independent of the
duration of the corresponding dimer.
The figure in the center shows full two chains.
The panel on the right shows an enlargement of the interfacial region.
} \label{photo}
\end{figure}

We have considered self-association of two $\alpha$-synuclein chains under
fairly dilute conditions as described in the Methods section. 
A dimer is considered to be formed if there is at least one contact
that connects the chains.
The C panel of Figure \ref{aggr} shows
that the residue-residue contacts that are the most engaged in coupling
two chains together, as assessed 
throughout the evolution,
are not those which drive the formation of transient secondary structures
in the individual chains. However, they are consistent with the
experimental findings discussed in the introduction. Figure \ref{photo}
shows an example of association in which 7 of the 10 top most
frequent contacts are present. They are indicated by the red lines
and they link the center parts of the chains.
The bottom-right panel of Figure \ref{aggr} shows the full contact map -- the
over-all pattern is seen to be distinct from the one obtained for the single-chain
implicit solvent calculation.\\

The top most frequent connecting contacts are shown in panel C of
Figure \ref{aggr}. They are seen to be mostly within the $h_2$ regions
of the two chains and then in the parts connecting $h_1$ with $h_2$.
There are also important contacts between $h_2$ and residue 111.\\

\begin{figure}[h]
\centering
\includegraphics[width=0.8\textwidth]{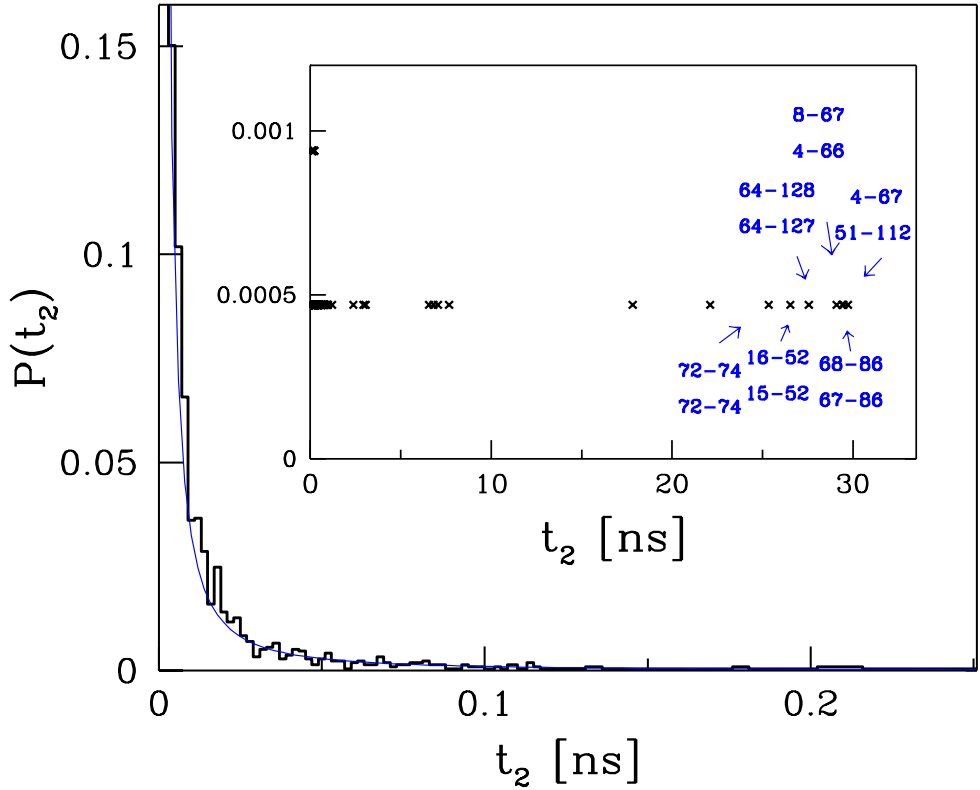}
\caption{The distribution of the duration times, $t_2$, of dimers (the
solid black line). 
The thinner blue line
corresponds to the power law fit  $(t_2/t_{2p})^{-3/2}$  with $t_{2p}$=0.002 ns.
It provides a better fit than exponential (not shown) with the
characteristic time scale of 0.005 ns.
The inset shows the same distribution in a different scale that is focused 
on the long-lasting and separate association events. Some of such events still
persisted at the end of the simulations. The numbers show the two longest lasting 
contacts corresponding to the particular long-lived events.
} \label{times}
\end{figure}

Figure \ref{times} shows the distribution of the duration times of the 
dimers. It is seen that most of the association events are short-lived: their 
life span does not exceed 0.2 ns. It seems that a power-law decay of the distribution
describes the behavior better than an exponential low (see the caption
of Figure \ref{times}). However, there are also events that last for tens of ns.
These correspond to the data points shown in the inset of the figure.
The corresponding most frequent contacts are different in each
event indicating existence of many pathways to associate. \\

It is interesting to point out that one of the  most frequent association
contacts is 4-67. Thus, if one of the chains is in the knotted state then
association involving residue 4 is expected to extend the duration 
of this state by making the knot deeper, similar to the effects of the procedures
described in ref. \cite{Bustamante}.\\

\section{Conclusions}

When analyzing the structural propensities, we have
found, not surprisingly, that the solvent increases the
disorder substantially, which shows as an enhancement in the C-content.
The types of the transient secondary structures that are detected
are the same indicating that for most computational purposes
the implicit-solvent approach is sufficient.\\

The nature of the solvent, however, may be important when assessing the
topological features. This appears to be the case of $\alpha$-synuclein.
This protein supports formation of knots in the explicit-solvent
case but not in the implicit-solvent one, or, at least,
there is a reduction in the probability of making a knot.
In any event, our results suggest that
the intrinsically disordered
proteins can generally support transient knots. 
It has been already demonstrated that sufficiently long poly-glutamine 
chains \cite{Angel}, which are also disordered, can support both
deeply- and shallowly-knotted conformations (the implicit-solvent calculation;
see also a coarse-grained study \cite{Mioduszewski}).
The presence of the knots in these chains has been suggested \cite{Angel1}
to be responsible for the monomeric-level toxicity leading to the Huntington
disease. The toxicity results from the fact that the knots may hinder or even jam the
degradation by the proteasomes (see a related discussion in ref.\cite{Bustamante})
and thus enhance the concentration of the chains in the cytosol.
In the case of $\alpha$-synuclein, the knots should enhance accumulation
of the proteins which, in turn, leads to an enhancement of multimeric aggregation.
It remains to be elucidated, however, whether the knotted states of $\alpha$-synuclein 
may lead to other physiologicallly relevant aspects.\\

We have already mentioned that the SAXS experiments \cite{Araki}
at the physiological pH yield the average $R_g$ of the human blood derived
$\alpha$-synuclein to be close to the value obtained by our explicit-solvent
calculations. While about 20\% of the durations of the explicit-solvent trajectories 
indicate the presence of the knots, the close agreement is not yet a proof
of their existence. It would be interesting to perform single-molecule
stretching experiments, similar to those done for structured proteins \cite{Rief},
to determine the fully stretched lengths. We expect the distributions of such
data to be two-peaked, with the shorter $L$ peak corresponding to the knotted
conformations.\\

\textbf{Acknowledgements}
We thank P. Robustelli for his inspiration to study knotting in $\alpha$-synuclein,
for providing us with the all-atom explicit solvent trajectories, and
for making Table \ref{table1}.
We appreciate very useful discussions with E. Go{\l}a\'s and B. R\'o\.zycki.
MC has received funding from the National Science Centre (NCN), Poland, 
under grant No. 2018/31/B/NZ1/00047.
This project is a part of the European COST Action EUTOPIA.

\clearpage

\begin{table}
\begin{center}
\caption{Comparison of calculated RMSD from experimental measurements for simulations of 
$\alpha$-synuclein. We compare the agreement of 30 $\mu$s explicit solvent trajectories 
of $\alpha$-synuclein from ref. \cite{Robustelli} run with c22*/TIP4P-D, 
c36M/TIP3P-CHARMM, a99SB­-disp, and a99SB*-ILDN/TIP3P and the knotted portions 
of the c22*/TIP4P-D and a99SB­-disp trajectories with previously reported experimental 
solution NMR measurements.  $R_g$ penalties were computed using an experimental value 
of 31.0 $\pm$ 5.0 from ref. \cite{Morar}. All classes (C$_\alpha$, H$_\alpha$, HN, C', 
C$_\beta$) of chemical shifts (CS) are reported in ppm; residual dipolar couplings
(RDCs) and indirect dipole--dipole couplings (J-couplings) are in Hz; $R_g$ is in {\AA};
paramagentic relaxation enhancements (PREs) and the scores are unitless.  
NMR observables and 
force field (FF) scores were calculated as previously reported \cite{Robustelli}, 
using only the subset of trajectories considered here.  
The $\textrm{Combined FF Score}$ is defined as $\left(\textrm{CS}_\textrm{Score}+\textrm{NMR}_\textrm{Score}\right)/{2}+\textrm{Rg}_\textrm{Penalty}$, $\textrm{Rg}_\textrm{Penalty}=\left(|\textrm{Rg}_\textrm{Exp}-\textrm{Rg}_\textrm{Sim}|-\textrm{Rg}_\textrm{Exp error}\right)/\textrm{Rg}_\textrm{Exp}$, 
where the $\textrm{CS}_\textrm{Score}$ is determined by normalizing the RMSD for 
each class of chemical shift by the smallest RMSD observed for the seven force fields and 
taking an average of the normalized RMSDs over all sets of experimental chemical shifts. 
The $\textrm{NMR}_\textrm{Score}$ is computed
analogously for all additional classes of NMR measurements and $\textrm{Rg}_\textrm{Exp error}$ 
is an experimentally estimated error of Rg.
We note that the knotted 
portions of the c22*/TIP4P-D and a99SB­-disp trajectories show marginally worse 
agreement with experimental measurements compared to their parent trajectories, 
but still show large improvements relative to simulations run with similar force 
fields and the TIP3P and TIP3P-CHARMM water models.}
\label{table1}
\begin{tabular}{ |c|c|c|c|c|c|c| }
 \hline
C$_\alpha$ CS &\;\; 0.43 \;\; & \;\; 0.51 \;\; & \;\; 0.61 \;\; & \;\; 0.51 \;\; & \;\; 0.59 \;\;  & \;\; 0.88 \;\;\\
H$_\alpha$ CS & 0.15 & 0.18 & 0.20 & 0.14 & 0.20 & 0.31\\
HN CS & 0.90 & 1.09 & 1.63 & 1.46 & 1.89 & 3.73\\
C' CS & 0.43 & 0.44 & 0.57 & 0.31 & 0.44 & 0.69\\
C$_\beta$ CS & 1.06 & 1.04 & 1.27 & 1.04 & 1.22 & 1.60\\
RDC (Q) & 0.47 & 0.59 & 0.64 & 0.41 & 0.52 & 0.93\\
Rg & 23.34 & 21.69 & 18.39 & 36.76 & 24.76 & 15.53\\
PRE & 0.18 & 0.22 & 0.32 & 0.17 & 0.23 & 0.39\\
Backbone ${}^3$J$_\textrm{HNHA}$ & 0.66 & 0.69 & 1.13 & 1.11 & 1.31 & 1.06\\
Backbone ${}^3$J$_\textrm{CC}$ & 0.15 & 0.17 & 0.30 & 0.18 & 0.24 & 0.46\\
CS$_\textrm{Score}$ & 1.10 & 1.22 & 1.54 & 1.16 & 1.50 & 2.43 \\
NMR$_\textrm{Score}$ & 1.06 & 1.23 & 1.81 & 1.23 & 1.56 & 2.33\\
Rg$_\textrm{Penalty}$ & 0.09 & 0.14 & 0.25 & 0.02 & 0.04 & 0.34\\
Combined FF Score & 1.16 & 1.37 & 1.92 & 1.22 & 1.57 & 2.72\\
 \hline
\end{tabular}
\end{center}
\end{table}

\clearpage

\clearpage


\end{document}